\documentclass[aip,pop,reprint,showpacs,superscriptaddress]{revtex4-1}

\usepackage{graphicx}
\usepackage{dcolumn}
\usepackage{bm}
\begin{document}

\title{Dynamics of ultra-intense circularly polarized solitons under inhomogeneous plasmas}

\author{Dong Wu}
\email{wudongphysics@gmail.com}
\affiliation{Center for Applied Physics and Technology, \\Peking University, Beijing, 100871, China.}
\affiliation{Key Laboratory of High Energy Density Physics Simulation, \\Ministry of Education,
Peking University, Beijing, 100871, China. }
\author{C. Y. Zheng}
\email{zheng\_chunyang@iapcm.ac.cn}
\affiliation{Center for Applied Physics and Technology, \\Peking University, Beijing, 100871, China.}
\affiliation{Key Laboratory of High Energy Density Physics Simulation, \\Ministry of Education,
Peking University, Beijing, 100871, China. }
\affiliation{Institute of Applied Physics and Computational Mathematics, \\Beijing, 100088, China.}
\author{X. T. He}
\email{xthe@iapcm.ac.cn}
\affiliation{Center for Applied Physics and Technology, \\Peking University, Beijing, 100871, China.}
\affiliation{Key Laboratory of High Energy Density Physics Simulation, \\Ministry of Education,
Peking University, Beijing, 100871, China. }
\affiliation{Institute of Applied Physics and Computational Mathematics, \\Beijing, 100088, China.}
\date{\today}
\begin{abstract}   
The dynamics of the ultra-intense circularly polarized solitons under inhomogeneous plasmas are examined. The interaction is modeled by the Maxwell and relativistic hydrodynamic equations and is solved with fully implicit energy-conserving numerical scheme. It is shown that a propagating weak soliton can be decreased and reflected by increasing plasma background, which is consistent with the existing studies based on hypothesis of weak density response. However it is found that ultra-intense soliton is well trapped and kept still when encountering increasing background. Probably, this founding can be applied for trapping and  amplifying high-intensity laser-fields.
   
\end{abstract}
\maketitle

The interaction between lasers and plasmas has attracted much attention since the invention of chirped pulsed amplification technology,
\cite{Opt.Commun.56.219,Phys.Fluids.B.4.2315}
which makes ultra-intense laser with intensity ($I_0 > 10^{18}$ W/cm$^2$) applicable.
Interaction between such intense lasers and plasmas exhibits a rich variety of interesting nonlinear phenomena.
One such phenomenon, namely the dynamics of electromagnetic solitons, has drawn much attention analytically, numerically and experimentally.
\cite{PhysRevLett.82.3440,PhyPla.20.033101,PhysRevA18.1591,PhysRevLett.69.1383,PhyPla.1.745,PhysRevLett.74.2002,Phys.Rev.E.55.1011,PhysRevLett.87.185004,
PhyPla.3.2693,PhysRevLett.76.3975,PhyPla.3.2693,Phys.Rev.E.54.1870,Phys.Rev.E.58.4937,Phys.Rev.E.65.016405,
Phys.Rev.E.70.036403,Phys.Rev.A.46.6608,PhysRevLett.87.275002,PhyPla.12.062308,PhyPla.13.092302,Phys.Rev.E.62.1234,
PhysRevLett.37.693,Phys.Rev.E.65.066406,Phys.Rev.E.58.4890,PhysRevLett.83.3434,
PhyPla.13.032309,PhysRevLett.68.3172,PhyPla.13.074504,PhyPla.13.032309,PhyPla.18.112112,
PhyPla.14.072307,Phys.Lett.A.377.473,plasma.phys.cont.fusion.47.A73}
As nearly $30\%\sim40\%$ of the laser pulse energy can be transformed into solitons.\cite{PhysRevLett.82.3440,PhyPla.20.033101}
This fairly high efficiency of electromagnetic energy transformation indicates that solitary
waves can play an important role in the development of the interaction between the laser
pulses and plasmas. In fact, many theoretical problems are still open even for one-dimensional solitons, namely the properties of linearly polarized solitons\cite{PhyPla.20.033101} and the soliton dynamics in the presence of plasma inhomogeneity\cite{PhysRevLett.37.693,Phys.Rev.E.65.066406,Phys.Rev.E.58.4890,Phys.Rev.E.58.4890,PhysRevLett.83.3434,
PhyPla.13.032309}. 
In reality, the background plasma density is always inhomogeneous,
thus it is of great necessity to investigate the dynamics of solitons under inhomogeneous plasma background.

Researches on the propagation of solitons under inhomogeneous background plasma have been conducted by various authors.\cite{PhysRevLett.37.693,Phys.Rev.E.65.066406,Phys.Rev.E.58.4890,PhysRevLett.83.3434,
PhyPla.13.032309}
The pioneering work in this field is on the theoretical studies of the Langmuir soliton through an inhomogeneous plasma.
Their analytical conclusions have demonstrated the acceleration (deceleration) of solitons in the presence of decreasing (increasing) plasma background.\cite{PhysRevLett.37.693} However there is a obviously shortcoming in this work, the group velocity of the soliton is treated as an independent parameter. In fact, the velocity, frequency and amplitude of the soliton are closely related, which is determined by an eigenvalue condition.\cite{PhysRevLett.68.3172}
Later on, the propagation of electromagnetic solitonic structures under inhomogeneous plasma has been investigated.\cite{Phys.Rev.E.65.066406,Phys.Rev.E.58.4890} 
However, all these related researches have been limited to the hypothesis of weak density response. \cite{PhysRevLett.68.3172}
The weak density response theory can not deal with ultra-intense solitons accompanied with heavy plasma density cavitation. 
The dynamic behavior of solitons under rare background plasma density has been studied by 2D Particle-in-Cell simulations,\cite{PhysRevLett.83.3434} which shows that the solitons are accelerated toward the vacuum-plasma interface with an acceleration proportional to the gradient of the
plasma density. 
Recently, fluid-Maxwell simulation has been carried out to systematically study the dynamic behavior of solitons under inhomogeneous plasma.\cite{PhyPla.13.032309} 
They have arrived at the conclusion$-$\textit{there is no qualitative difference in the propagation features of large-amplitude solutions and the small-amplitude solutions of the nonlinear Schroedinger equation soliton variety, through the inhomogeneous plasma background}.
They have found that both weak and strong solitons can be decreased and reflected by the increasing plasma background.
In their simulations, the propagating soliton is put in as an initial condition, where the amplitude of the strong soliton is $a\sim0.9$ ($a=eE/m_ec\omega_0$) and that of weak soliton is $a\sim0.4$. What about the ultra-intense soliton with amplitude $a>1.0$, can its behavior still be the same?

In this paper, the dynamic behavior of ultra-intense solitons under inhomogeneous plasma background has been conducted by fluid-Maxwell simulation which is solved with fully implicit energy-conserving numerical scheme. 
It is shown that the weak propagating soliton can be decreased and reflected by the increasing plasma background, which is consistent with the existing studies and can be described by the weak-density response theory. 
However the ultra-intense electromagnetic soliton can not be decreased and reflected by the increasing background, but is trapped and kept still in the plasma density cavitation.  

Compared with existing studies, 
we address the issue of propagation of electromagnetic soliton under inhomogeneous background in a more general context. 
(a) The hypothesis of weak density response is dropped, and we can investigate the dynamics of ultra-intense solitons.
(b) The background plasma is not restricted to be of rare density, instead, over-dense plasma background is considered.
(c) The soliton is self-generated by the interaction between laser and plasma as a more general process, which is not put in as an initial condition.

The governing equations are the relativistic hydrodynamic fluid equations coupled with the full system of Maxwell equations. 
As the electron quiver velocity is much greater than the thermal velocity in relativistic intense laser light, 
we treat the plasmas as cold liquid. The normalized full system of equations are shown below\cite{PhyPla.20.033101,PhyPla.12.062308,Phys.Rev.E.62.1234}
\begin{eqnarray}\label{1,2,3,4,5,6,7}
&& \frac{\partial^2 a_{y,z}}{\partial t^2}-\frac{\partial^2 a_{y,z}}{\partial x^2}+\left(\frac{n_e}{\gamma_e}+Z^2\frac{n_i}{m\gamma_i}\right)a_{y,z}=0, \\
&& \frac{\partial n_{e,i}}{\partial t}+\frac{\partial}{\partial x}\left(\frac{n_{e,i} p_{e,i}}{\gamma_{e,i}}\right)=0, \\
&& \frac{\partial p_e}{\partial t}+\frac{\partial \gamma_e}{\partial x}-\frac{\partial \phi}{\partial x}=0, \\
&& \frac{\partial p_i}{\partial t}+\frac{\partial \gamma_i}{\partial x}+\frac{Z}{m}\frac{\partial \phi}{\partial x}=0, \\
&& \frac{\partial^2 \phi}{\partial x^2}=\left(n_e-n_i\right), \\
&& \gamma_e=\left[1+a_y^2+a_z^2+p_e^2\right]^{1/2}, \\
&& \gamma_i=\left[1+\left(a_y/m\right)^2+\left(a_z/m\right)^2+p_i^2\right]^{1/2},
\end{eqnarray}
where $t=\omega_{pe}t$, $x=\omega_{pe}x/c$, $a_{y,z}=eA_{y,z}/m_ec^2$, $n_{e,i}=n_{e,i}/n_0$, $p_{e,i}=p_{e,i}/m_ec$, and $\phi=e\phi/m_ec^2$ are the normalized time, space, wave vector potential, electron and ion density and momentum (along the laser propagation direction $x$), and electrostatic potential of the space charge field, respectively. Furthermore, $m$ is the ion/electron mass ratio, and $\omega_{pe}$ and $n_0=m_e\omega_{pe}^2/4{\pi}e^2$ are the plasma frequency and its related plasma density.

To solve the equations, we use a fully implicit energy-conserving numerical scheme.\cite{PhyPla.20.033101,numerical} 
The boundary conditions are $a_{y,z}=0$, $\phi=0$ and $n_{e,i}=0$ at both sides of the simulation box. Our model is a global simulation, namely, at initial time, the incident laser and plasma are apart form each other. The laser impacts the plasma, and the evolution of vacuum-plasma interface is demonstrated in detail.\cite{PhyPla.20.033101}
Unlike the existing studies that the propagating soliton is put in as an initial condition, 
in our simulation the soliton is self-generated through the interaction between laser and plasma.  
It is generally accepted that at vacuum region the fluid treatment of plasma is invalid, because the very definition of a plasma calls for an accumulation of macroscopic number of particles. Numerical integrating of Eqs. (4) and (5) result in unphysical results of $p_{e,i}$ at the zero plasma density region. In our simulation, $p_{e,i}$ is restricted to be zero at the vacuum region. As the zero plasma density leads to zero current density $J_x=0$, it has no effect on the solution of coupled Maxwell Eq. (1), which makes the global fluid-Maxwell simulation applicable.
\cite{PhyPla.20.033101,PhyPla.12.062308}   

The initially incident laser is of the form,
\begin{eqnarray}\label{8}
&& a_y=a_m\cos^2[(x-x_{p0})\pi/2\tau]\cos[\omega_0(t-x)], \nonumber \\
&& a_z=a_m\cos^2[(x-x_{p0})\pi/2\tau]\sin[\omega_0(t-x)], 
\end{eqnarray}
where $x_p$ is the position of peak laser intensity, $x_p-\tau<x<x_p+\tau$, 
$\tau=5.0\lambda_0$ ($\lambda_0=2\pi/\omega_0$) is the laser duration, $\omega_0=0.88$ is the laser frequency and $a_m=0.71$ is the maximum laser amplitude. 
The initial plasma density distribution is of the form,
\begin{eqnarray}\label{9}
&& n_e=n_p\sin^2[(x-x_{p1})\pi/2(x_{p2}-x_{p1})], \nonumber \\
&& x_{p1} \leq x \leq x_{p2}; \nonumber \\
&& n_e=n_p, x_{p2} < x \leq x_{p3}; \nonumber \\
&& n_e=n_p+n_p\alpha\sin^2[(x-x_{p3})\pi/2(x_{p4}-x_{p3})], \nonumber \\ 
&& x_{p3} < x \leq x_{p4}; \nonumber \\
&& n_e=n_p(1+\alpha)\cos^2[(x-x_{p4})\pi/2(x_{p5}-x_{p4})], \nonumber \\
&& x_{p4} < x \leq x_{p5}. 
\end{eqnarray}
To describe an increasing plasma background, we have $n_p=1.0$ ($=1.29n_c$), $\alpha=0.2$, $x_{p1}=0.0\lambda_0$, $x_{p3}=4.0\lambda_0$ and 
$x_{p4}=8.0\lambda_0$. The steepness factor is define as $s=x_{p2}-x_{p1}$, which plays an important role in the laser plasma interaction.  
\begin{figure}
\includegraphics[width=8.5cm]{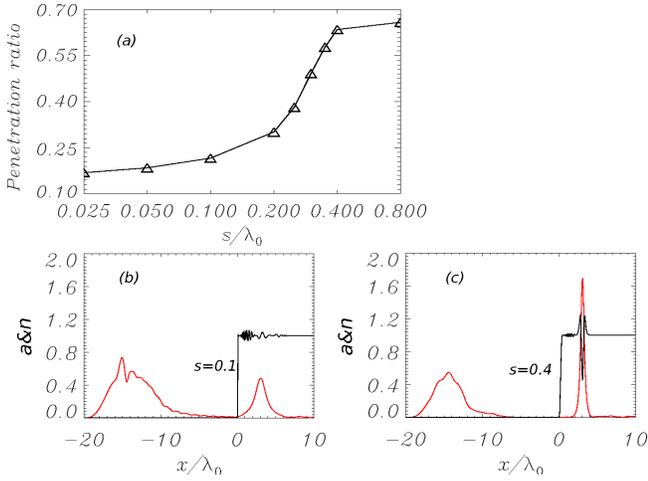}
\caption{\label{fig1} (color online) (a) The penetration ratio of the incident laser pulse vs. shape factor $s$.
(b) and (c) The reflected and penetrated laser structure for different vacuum-plasma interface steepness, 
$s=0.1\lambda_0$ and $s=0.4\lambda_0$, respectively. 
The red curve is for the normalized laser amplitude $a=(a_y^2+a_z^2)^{1/2}$ and black curve is for the plasma electron density.
Here the simulation parameters are $a_y=a_z=0.71$, $\omega_0=0.88$, $\tau=5.0\lambda_0$, $x_{p0}=-5.02\lambda_0$ and $x_{p1}=0.0\lambda_0$.}
\end{figure}
As shown in Fig.\ \ref{fig1}(a), the penetration ratio is closely related to $s$, the steepness of the vacuum-plasma interface. 
The penetration ratio is defined as  $(\epsilon_{in}-\epsilon_{bk})/\epsilon_{in}$, 
where $\epsilon_{in}$ is the initial laser energy and $\epsilon_{bk}$ is the reflected light energy.
We can see that a rather smooth vacuum-plasma interface allows for a strong penetration of laser pulse.
According to this relation, we can control the amplitude of the soliton by adjusting the steepness of vacuum-plasma interface.  
For a rather steep vacuum-plasma interface, as shown in Fig.\ \ref{fig1}(b), most of the incident laser is reflected and the penetrated electromagnetic soliton is weak. 
However for a rather smooth vacuum-plasma interface, as shown in Fig.\ \ref{fig1}(c),
the penetrated electromagnetic soliton is ultra-intense with amplitude $a\sim1.7$. 
The behavior of ultra-intense soliton is beyond the prediction of weak-density response theory, and it is of great interest to numerically study its dynamics under inhomogeneous plasma background.

\begin{figure}
\includegraphics[width=8.5cm]{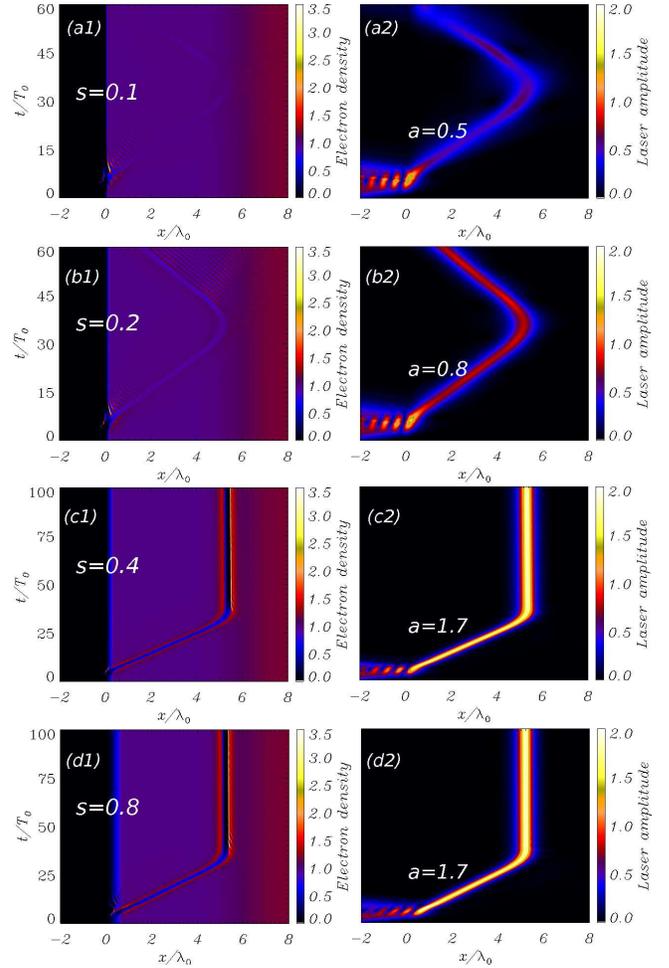}
\caption{\label{fig2} (color online) The space-time evolution of plasma density and laser amplitude under increasing plasma background,
where the column\ (1) is for plasma density and column\ (2) for laser amplitude.
Different cases of vacuum-plasma steepness (or soliton amplitude) with $s=0.1\lambda_0$, $s=0.2\lambda_0$, $s=0.4\lambda_0$ and $s=0.8\lambda_0$ (or $a=0.5$, $a=0.8$, $a=1.7$ and $a=1.7$) are shown in (a), (b), (c) and (d), respectively.
Here the simulation parameters are $a_y=a_z=0.71$, $\omega_0=0.88$, $\tau=5.0\lambda_0$, $x_{p0}=-5.02\lambda_0$, $x_{p1}=0.0\lambda_0$,
$x_{p3}=4.0\lambda_0$, $x_{p4}=8.0\lambda_0$, $n_p=1.0$ and $\alpha=0.2$.}
\end{figure}

Figure\ \ref{fig2} shows the space-time evolution of plasma density (left column) and laser amplitude (right column) under inhomogeneous background.
For the small-amplitude soliton with $a\sim0.5$, as clearly demonstrated in Fig.\ \ref{fig2} (a1), the plasma density is almost undisturbed.
Weak density response theory can well deal with the behavior of such small-amplitude soliton under inhomogeneous plasma background.
The soliton undertakes deceleration and eventual reflection when encountering increasing plasma background, as shown in Fig.\ \ref{fig2} (a2).
As for the more intense soliton with $a\sim0.8$, the plasma density is more or less distributed and
a shallow cavitation is formed moving with the propagating soliton, which is shown in  Fig.\ \ref{fig2} (b1).
Strictly speaking, the weak density response treatment is no longer proper to describe the dynamic behavior of soliton with $a\sim0.8$, however its behavior, deceleration and reflection, is almost the same as the small-amplitude soliton. 
It seems that \textit{there is no qualitative difference in the propagation features of large-amplitude solutions and the small-amplitude solutions of the nonlinear Schroedinger equation soliton variety, through the inhomogeneous plasma background}.
However, Fig.\ \ref{fig2} (c) and (d) indicate that large-amplitude solitons with $a\sim1.7$ behave dramatically different from that of small-amplitude. The plasma density is strongly affected by the ponderomotive force of large-amplitude soliton, 
and a deep density cavitation is formed. Unlike the behavior of small-amplitude soliton, the large-amplitude soliton firstly experiences deceleration, then is well trapped and kept still in the deep plasma cavitation, when encountering increasing plasma background . 

For a general wave packet with group velocity $\beta$ propagating in the plasma, we have $\beta=(1-\omega_{pe}^2/\omega^2)^{1/2}=(1-n_p/n_{e})^{1/2}$,\cite{PhyPla.13.032309} where $n_p=1.0$. This relation demonstrates that the group velocity is closely related to the plasma background density.
At $n_{zero}=1/(1-\beta^2)=1/(1-0.17^2)=1.03$, the group velocity of soliton is decreased to be zero.
For small-amplitude soliton with almost no density response, the group velocity of soliton is totally determined by the plasma background density profile. However for large-amplitude soliton, such as Fig.\ \ref{fig1}(c), the density of the cavitation edge can be as high as $1.2$, which is even higher than the increasing plasma background  density. The high-amplitude soliton is surrounded by high plasma density located at its both front and back sides, which are both higher than $n_{zero}$. In this case, the high-amplitude soliton can not be reflected, as the small-amplitude soliton does, by the increasing plasma background density. 
This high-amplitude and well trapped soliton can be the basis for trapping and amplifying high-intensity laser-fields between two solid density foils.\cite{PhysRevLett.89.275004}

\begin{figure}
\includegraphics[width=8.5cm]{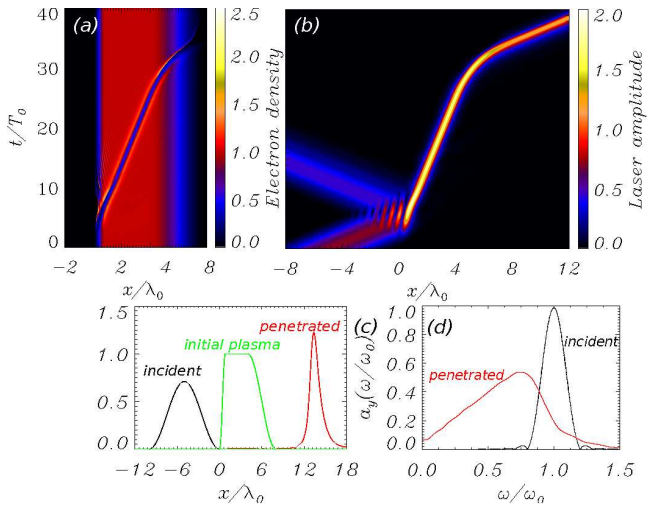}
\caption{\label{fig3} (color online) (a) and (b) The space-time evolution of plasma density and laser amplitude under decreasing plasma background.
(c) The back curve represents the incident laser, green curve represents the initial plasma density profile, and red curve for the penetrated laser pulse. 
(d) The spectral intensity of the incident laser and penetrated laser, where the black curve is for incident laser and
red curve is for penetrated laser.
Here the simulation parameters are $a_y=a_z=0.71$, $\omega_0=0.88$, $\tau=5.0\lambda_0$, $x_{p0}=-5.02\lambda_0$, $x_{p1}=0.0\lambda_0$,
$x_{p3}=x_{p4}=4.0\lambda_0$, $x_{p5}=8.0\lambda_0$, $n_p=1.0$ and $s=0.8\lambda_0$.}
\end{figure}

Figure\ \ref{fig3}(a) and (b) show the space-time evolution of plasma density and laser amplitude under decreasing plasma background.
As expected, the soliton undertakes acceleration through the decreasing background. The electromagnetic soliton penetrates and enters the backside vacuum of the plasma, where no second reflection occurs on the backside vacuum-plasma interface. As can be seen from Fig.\ \ref{fig3}(c), the amplitude and duration of the penetrated laser are stronger and shorter than that of the initially incident laser. Figure\ \ref{fig3}(d) shows the spectrum distributions of the incident and penetrated laser, which indicates that the penetrated laser is a good source of low-frequency burst of electromagnetic radiation. 
 
In summary, the dynamic behavior of ultra-intense soliton under inhomogeneous plasma background has been conducted by fluid-Maxwell simulation which is solved with fully implicit energy-conserving numerical scheme. 
The amplitude of propagating soliton can be well controlled by the adjusting the steepness of vacuum-plasma interface.
It is shown that the small-amplitude soliton can be decreased and reflected by the increasing plasma background.
However the large-amplitude soliton is well trapped and kept still encountering increasing plasma background. 
Probably, this founding can be applied for trapping and amplifying high-intensity laser-fields.
The amplitude and duration of the penetrated laser pulse are much stronger and shorter than that of the initially incident laser. The penetrated laser is a good source of low-frequency burst of electromagnetic radiation.

\begin{acknowledgments}
This work was supported by the National Natural Science Foundation of China (Grant Nos. 11075025), 
the Ministry of Science and Technology of China (Grant No. 2011GB105000) and National Basic Research Program of China (grant No. 2013CB834100).
\end{acknowledgments}

{}

\begin{thebibliography}{99}
\bibitem{Opt.Commun.56.219} D. Strickland and G. Mourou, Opt. Commun. 56, 219 (1985).
\bibitem{Phys.Fluids.B.4.2315} G. Mourou and  D. Umstadter, Phys. Fluids B 4, 2315 (1992).
\bibitem{PhysRevLett.82.3440} S. V. Bulanov, T. Zh. Esirkepov, N. M. Naumova, F. Pegoraro and V. A. Vshivkov, Phys. Rev. Lett. 82, 3440 (1999).
\bibitem{PhyPla.20.033101} Dong Wu, C. Y. Zheng, X. Q. Yan, M. Y. Yu and X. T. He, Phys. Plasmas 20, 033101 (2013).
\bibitem{PhysRevA18.1591} M. Y. Yu, P. K. Shukla, and K. H. Spatsheck, Phys. Rev. A 18, 1591 (1978).
\bibitem{PhysRevLett.69.1383} S. C. Wilks, W. L. Kruer, M. Tabak, and A. B. Langdon, Phys. Rev. Lett. 69, 1383 (1992).
\bibitem{PhyPla.1.745} S. V. Bulanov, N. M. Naumova, and F. Pegoraro, Phys. Plasma 1, 745 (1994).
\bibitem{PhysRevLett.74.2002} E. Lefebvre and G. Bonnaud, Phys. Rev. Lett. 74, 2002 (1995).
\bibitem{Phys.Rev.E.55.1011} E. Lefebvre and G. Bonnaud, Phys. Rev. E 55, 1011 (1997).
\bibitem{PhysRevLett.87.185004} N. M. Naumova , S. V. Bulanov, T. Zh. Esirkepov, D. Farina, K. Nishihara, F. Pegoraro, H. Ruhl and A. S. Sakharov, Phys. Rev. Lett. 87, 185004 (2001).
\bibitem{PhysRevLett.76.3975} A. Pukhov and J. Meyer-ter-Vehn, Phys. Rev. Lett. 76, 3975 (1996).
\bibitem{PhyPla.3.2693} S. Guerin, P. Mora, J. C. Adam, A. Heron, and G. Laval, Phys. Plasmas 3, 2693 (1996).
\bibitem{Phys.Rev.E.54.1870} H. Sakagami and K. Mima, Phys. Rev. E 54, 1870 (1996).
\bibitem{Phys.Rev.E.58.4937} K. Nagashima, Y. Kishimoto, and H. Takuma, Phys. Rev. E 58, 4937 (1998).
\bibitem{Phys.Rev.E.65.016405} B. Shen and J. Meyer-ter-Vehn, Phys. Rev. E 65, 016405 (2001).
\bibitem{Phys.Rev.E.70.036403} Baifei Shen, M. Y. Yu, and Ruxin Li, Phys. Rev. E 70, 036403 (2004).
\bibitem{Phys.Rev.A.46.6608} V. I. Berezhiani, D. D. Tskhakaya, and P. K. Shukla, Phys. Rev. A 46, 6608 (1992).
\bibitem{PhysRevLett.87.275002} M. Tushentsov, A. Kim, F. Cattani, D. Anderson, and M. Lisak, Phys. Rev. Lett. 87, 275002 (2001).
\bibitem{PhyPla.12.062308} V. I. Berezhiani, D. P. Garuchava, S. V. Mikeladze, K. I. Sigua, and N. L. Tsintsadze, Phys. Plasmas 12, 062308 (2005).
\bibitem{PhyPla.13.092302} G. Lehmann, E. W. Laedke, and K. H. Spatschek, Phys. Plasmas 13, 092302 (2006).
\bibitem{Phys.Rev.E.62.1234} F. Cattani, A. Kim, D. Anderson and M. Lisak, Phys. Rev. E 62, 1234 (2000).
\bibitem{PhysRevLett.37.693} H. H. Chen and C. S. Liu, Phys. Rev. Lett. 37, 693 (1976).
\bibitem{Phys.Rev.E.65.066406} M. R. Rouhani, H. Abbasi, H. H. Pajouh, P. K. Shukla, and N. L. Tsintsadze, Phys. Rev. E 65, 066406 (2002).
\bibitem{Phys.Rev.E.58.4890} N. L. Tsintsadze, J. T. Mendonca and L. Oliveira e Silva, Phys. Rev. E 58, 4890 (1998).
\bibitem{PhysRevLett.83.3434} Y. Sentoku, T. Zh. Esirkepov, K. Mima, K. Nishihara, F. Califano, F. Pegoraro, H. Sakagami, Y. Kitagawa,
N. M. Naumova and S. V. Bulanov, Phys. Rev. Lett. 83, 3434 (1999).
\bibitem{PhyPla.13.032309} Vikrant Saxena, Amita Das, Abhijit Sen and Predhiman Kaw, Phys. Plasmas 13, 032309 (2006).
\bibitem{PhysRevLett.68.3172} P. K. Kaw, A. Sen, and T. Katsouleas, Phys. Rev. Lett. 68, 3172 (1992).
\bibitem{PhyPla.13.074504} Bai-Song Xie and Shu-Cheng Du Phys. Plasmas 13, 074504 (2006).
\bibitem{PhyPla.18.112112} Sita Sundar, Amita Das, Vikrant Saxena, Predhiman Kaw and Abhijit Sen, Phys. Plasmas 18, 112112 (2011).
\bibitem{PhyPla.14.072307} Vikrant Saxena, Amita Das, Sudip Sengupta, Predhiman Kaw and Abhijit Sen, Phys. Plasmas 14, 072307 (2007)
\bibitem{Phys.Lett.A.377.473} V. Saxena, I. Kourakis, G. Sanchez-Arriaga and E. Siminos, Phys. Lett. A 377, 473 (2013).
\bibitem{plasma.phys.cont.fusion.47.A73} Daniela Farina and Sergei V Bulanov, Plasma Phys. Control. Fusion 47, A73 (2005).
\bibitem{numerical} The details of the numerical scheme shall be presented elsewhere.
\bibitem{PhysRevLett.89.275004} B. Shen and M. Y. Yu, Phys. Rev. Lett. 89, 275004 (2002).
\end{thebibliography}
\end{document}